\documentclass[11pt]{article}
\usepackage{times}
\usepackage{sidecap}
\usepackage{geometry}
\usepackage{floatrow}
\usepackage{float}
\usepackage{hyperref}
\usepackage{graphics,graphicx}
\usepackage{subcaption}
\usepackage{epsf,wrapfig}

\usepackage{natbib,bm}

\usepackage{paralist}
\let\olditem\item
\renewenvironment{thebibliography}[1]{%
  \section*{\refname}
  \let\par\relax
  \renewcommand{\item}[1][]{\olditem[\textbullet]}%
  \inparaenum}{\endinparaenum}

\newcounter{aff}

\DeclareGraphicsExtensions{.jpg,.pdf,.png,.eps}

\usepackage{caption}
\usepackage{xcolor}

\usepackage{tcolorbox}
\usepackage{rotating}
\usepackage{graphicx}
\bibpunct{(}{)}{;}{a}{}{,}

\usepackage{soul}

\definecolor{darkcyan}{rgb}{0.0, 0.0, 0.7}

\definecolor{boxcolor}{RGB}{193, 229, 255}
\definecolor{figboxcolor}{RGB}{193, 229, 255}
\definecolor{boxcolorgray}{RGB}{227, 227, 227}
\definecolor{darkcyan}{rgb}{0.0, 0.0, 0.7}

\sethlcolor{boxcolorgray}

\def\gtrsim{\lower 2pt \hbox{$\, \buildrel {\scriptstyle >}\over
{\scriptstyle \sim}\,$}}
\def\lesssim{\lower 2pt \hbox{$\, \buildrel {\scriptstyle <}\over
{\scriptstyle \sim}\,$}}

\def\approxlt{\lower.2em\hbox{$\buildrel < \over \sim$}}
\def\approxgt{\lower.2em\hbox{$\buildrel > \over \sim$}}

\def\arcmin{$^{\prime}$}
\def\arcsec{$^{\prime\prime}$}

\def\be{\begin{equation}}
\def\ee{\end{equation}}
\def\bea{\begin{eqnarray*}}
\def\eea{\end{eqnarray*}}

\def\arcsec{^{\prime\prime}}
\def\arcmin{^\prime}

\def\jref@jnl#1{{\rm#1}}
\def\aj{\jref@jnl{AJ}}			
\def\araa{\jref@jnl{ARA\&A}}		
\def\apj{\jref@jnl{ApJ}}			
\def\apjl{\jref@jnl{ApJ}}		
\def\apjs{\jref@jnl{ApJS}}		
\def\ao{\jref@jnl{Appl.~Opt.}}		
\def\apss{\jref@jnl{Ap\&SS}}		
\def\aap{\jref@jnl{A\&A}}		
\def\aapr{\jref@jnl{A\&A~Rev.}}		
\def\aaps{\jref@jnl{A\&AS}}		
\def\azh{\jref@jnl{AZh}}			
\def\baas{\jref@jnl{BAAS}}		
\def\jrasc{\jref@jnl{JRASC}}		
\def\memras{\jref@jnl{MmRAS}}		
\def\mnras{\jref@jnl{MNRAS}}		
\def\pra{\jref@jnl{Phys.~Rev.~A}}	
\def\prb{\jref@jnl{Phys.~Rev.~B}}	
\def\prc{\jref@jnl{Phys.~Rev.~C}}	
\def\prd{\jref@jnl{Phys.~Rev.~D}}	
\def\pre{\jref@jnl{Phys.~Rev.~E}}	
\def\prl{\jref@jnl{Phys.~Rev.~Lett.}}	
\def\pasp{\jref@jnl{PASP}}		
\def\pasj{\jref@jnl{PASJ}}		
\def\qjras{\jref@jnl{QJRAS}}		
\def\skytel{\jref@jnl{S\&T}}		
\def\solphys{\jref@jnl{Sol.~Phys.}}	
\def\sovast{\jref@jnl{Soviet~Ast.}}	
\def\ssr{\jref@jnl{Space~Sci.~Rev.}}	
\def\zap{\jref@jnl{ZAp}}			
\def\nat{\jref@jnl{Nature}}		
\def\iaucirc{\jref@jnl{IAU~Circ.}}
\def\aplett{\jref@jnl{Astrophys.~Lett.}}
\def\apspr{\jref@jnl{Astrophys.~Space~Phys.~Res.}}
\def\bain{\jref@jnl{Bull.~Astron.~Inst.~Netherlands}}
\def\fcp{\jref@jnl{Fund.~Cosmic~Phys.}}
\def\gca{\jref@jnl{Geochim.~Cosmochim.~Acta}}
\def\grl{\jref@jnl{Geophys.~Res.~Lett.}}
\def\jcp{\jref@jnl{J.~Chem.~Phys.}}	
\def\jgr{\jref@jnl{J.~Geophys.~Res.}}	
\def\jqsrt{\jref@jnl{J.~Quant.~Spec.~Radiat.~Transf.}}
\def\memsai{\jref@jnl{Mem.~Soc.~Astron.~Italiana}}
\def\nphysa{\jref@jnl{Nucl.~Phys.~A}}
\def\physrep{\jref@jnl{Phys.~Rep.}}
\def\physscr{\jref@jnl{Phys.~Scr}}
\def\planss{\jref@jnl{Planet.~Space~Sci.}}
\def\procspie{\jref@jnl{Proc.~SPIE}}

\def\arcmin{\hbox{$^\prime$}}
\def\arcsec{\hbox{$^{\prime\prime}$}}

\def\ion#1#2{#1$\;${\small\rm\@Roman{#2}}\relax}

\newcount\lecurrentfam
\def\LaTeX{\lecurrentfam=\the\fam \leavevmode L\raise.42ex
\hbox{$\fam\lecurrentfam\scriptstyle\kern-.3em A$}\kern-.15em\TeX}
\def\sizrpt{
(\fontname\the\font): em=\the\fontdimen6\font, ex=\the\fontdimen5\font
\typeout{
(\fontname\the\font): em=\the\fontdimen6\font, ex=\the\fontdimen5\font
}}
\definecolor{dark}{rgb}{ 0.34, 0.09, 0.49}

\usepackage{fancyhdr,url}
\usepackage{transparent}
\usepackage{tablefootnote}
\usepackage{mathpazo}

\begin{document}

\noindent
\begin{raggedright}
{\huge\bfseries Planet forming disk populations and their connection to the environment  \\[0.3em]
\smallskip 
\large \textit{How do planets form in the most common star-forming conditions in the Galaxy? }
}\end{raggedright} 

\medskip 
\noindent
Thomas J. Haworth$^{1,*}$, Giovanni Rosotti$^2$, Carlo Manara$^3$, Richard Alexander$^4$, Rossella Anania$^5$, Til Birnstiel$^{6}$, Ryan Boyden$^{7}$,  Nicolás Cuello$^9$, Ian Czekala$^{19}$, Vardan Elbakyan$^{10}$, Stefano Facchini$^2$,  Fabiola Gerosa$^{11}$, Aashish Gupta$^{8}$,  Dominika Itrich$^{12}$, Luke Keyte$^1$, \'Agnes K\'osp\'al$^{20}$, Giuseppe Lodato$^2$, Enrique Macias$^{3}$,  Luca Matra$^5$, Rebecca Nealon$^{13}$, Sasha Paine$^{1}$, Danae Polychroni$^{14}$, Tyger Peake$^1$, Daniel Price$^{13}$, Lin Qiao$^1$, Álvaro Ribas$^1$, Sarah Sadavoy$^{21}$, 
Antranik A. Sefilian$^{12}$, Maxime Tanious$^{22}$,
Sierk van Terwisga$^{15}$, Claudia Toci$^{16}$, Diego Turrini $^{14}$, Eduard Vorobyov$^{17}$, Catherine Walsh$^{18}$, Andrew Winter$^1$, 
 Alice S. Booth$^{23}$, Liton Majumdar$^{24,25}$, Jenny Frediani$^{26}$ \\
\medskip
{
\footnotesize{
\noindent $^*$ t.haworth@qmul.ac.uk \\ \noindent Affiliations are listed at the end of the text
}
}

\begin{tcolorbox}[boxsep=3pt,left=0pt,right=0pt,top=1pt,bottom=1pt,colframe=boxcolor,colback=boxcolor]
\justify
\textbf{Abstract:} 
Existing sub-mm observatories such as ALMA have provided statistics on the planet-forming mass reservoirs within circumstellar disks around stars in stellar clusters out to distances of $\sim$~400~pc. However, these disks are not where most planets form. Many stars, and hence planets, form in more massive clusters, which are only found at larger distances. Environmental conditions that affect disks -- such as UV radiation field, ambient gaseous mass reservoir, and stellar density -- change significantly as a function of cluster mass, implying that our current knowledge of the disk population from current facilities is severely biased. Characterising planet formation in stellar clusters towards the peak of the cluster mass function requires determining the distribution of planet forming disk masses out to $\sim$3kpc, which encapsulates the nearby parts of the Sagittarius arm of the Milky Way and contains multiple massive stellar clusters. ALMA2040 will provide the leap in sensitivity needed to move beyond detections of only a handful of disks and robustly measure the slope and median of the disk mass distribution at these distances. This will mark the transition from studying individual detections to population-level studies of planet formation across diverse cluster environments. This has synergy with existing and upcoming measurements of accretion rates in these distant clusters, which combined provide a measure of the accretion timescale, enabling us to distinguish various evolutionary pathways and assess how they depend on cluster mass and star-forming environment.
\end{tcolorbox}

\newpage{}

\vspace{-0.7cm}
\section{Scientific context and open questions}

To understand exoplanet diversity \citep{2025arXiv250509754V} we have to understand how planets form within circumstellar disks, and hence how the disks themselves evolve over time. ALMA has enabled the study of planet forming disks in nearby star-forming regions at unprecedented levels of detail. Its other key impact has been providing statistical surveys of disk properties in star-forming regions out to approximately 400\,pc down to dust masses of $\gtrsim0.1$ Earth masses \citep[e.g.][]{2016ApJ...828...46A, 2017AJ....153..240A, 2018ApJ...860...77E}. This has enabled us to study the demographics and time evolution of key parameters, such as the disk dust masses -- which quantify a key part of the planet forming mass reservoir, resulting in some of the most highly cited papers produced using ALMA \citep[e.g.][]{2016ApJ...828...46A}.

\begin{figure}[!t]
    \centering
    \vspace{-0.4cm}
    \includegraphics[width=0.97\linewidth]{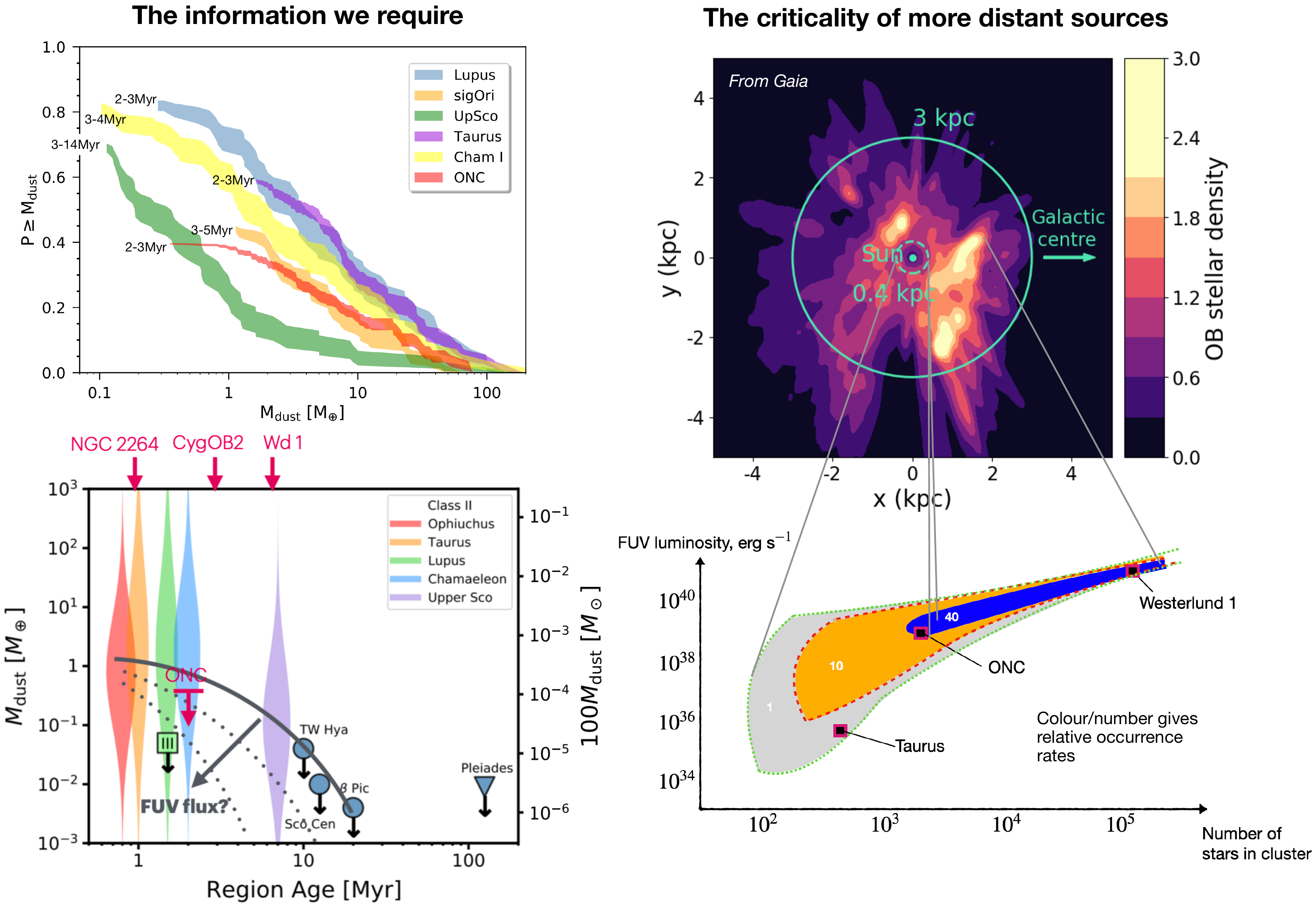}
    \caption{Current facilities (e.g. ALMA) have delivered key information on the disk mass distribution in different clusters and its variation over time \protect\citep[left panels,][]{2018ApJ...860...77E, 2023ASPC..534..539M}. These data hint at a possible sensitivity to environment (the ONC shows systematically lower masses, upper left). However, these datasets are drawn from within 0.4\,kpc of the Sun, as represented by the inner circle in the upper right panel showing the Gaia \protect\citep{2016A&A...595A...1G} OB star density centred on the Sun. Within that 0.4\,kpc distance, clusters are predominantly low mass, the external FUV luminosity is weak, whereas many stars will form in more massive clusters \citep[lower right panel, adapted from][]{2022EPJP..137.1132W}. We require observations out to 3\,kpc to capture the massive ($10^{4-5}$ star hosting) clusters where most stars, and thus planetary systems, form to determine how the disk dust masses evolve in such environments. }
    \label{fig:mainScience}
\end{figure}

However, there is still a critical shortfall in our understanding in that ALMA is only capable of sampling disks in stellar clusters with masses  two orders of magnitude below the peak of the distribution in terms of where stars form. This matters because the \textit{star-forming environment} in massive clusters is likely a major, perhaps even dominant, driver of disk evolution. For instance, strong far-ultraviolet (FUV) radiation fields are capable of driving photoevaporative winds \citep{2022EPJP..137.1132W, 2025OJAp....8E..54A}, the high stellar density and multiplicity can lead to gravitational perturbations and disk truncation \citep[e.g.][]{2023EPJP..138...11C}, and the high density of the star-forming gas followed by rapid dispersal via feedback can lead to a very different ambient mass reservoir for accretion onto star/disks over time compared to lower mass clusters \citep{2020NatAs...4.1158P, 2024A&A...683A.133G, 2026NatAs..10.1179P}. There is evidence that these effects do impact the bulk properties of planet forming disks: \cite{2018ApJ...860...77E} showed that disk masses in the Orion Nebula Cluster (ONC) are systematically lower, and the cumulative distribution of masses shallower, than those in lower mass clusters of similar age (upper left panel of Figure \ref{fig:mainScience}).  Relatedly, theoretical works suggest that this environmentally driven depletion of disks could significantly alter the resulting planetary system architectures \citep{2023MNRAS.522.1939Q,2026MNRAS.546ag034Q,  2025ApJ...979..120H}.

These considerations lead to  key open questions such as: 
\begin{enumerate}
    \item How does the global distribution of planet-forming disk masses evolve with time and vary with host-cluster properties, such as the cluster stellar mass, stellar density, UV luminosity, and molecular gas content (a source of accretion and shielding from irradiation)? 
    \item How do we unambiguously distinguish age effects and environmental effects? E.g. there is evidence that the $M_{\textrm{dust}}-M_{*}$ relationship possibly steepens with age \citep{2016ApJ...831..125P}.
    \item What fraction of disks are affected by external irradiation early enough ($<1$\,Myr) to impact planet formation?    
    \item How does the transport of material through disks depend on environmental effects, including irradiation, accretion from the interstellar medium (infall), dynamical encounters and multiplicity?
    \item In massive clusters there is a larger molecular cloud gas mass reservoir but also stronger massive star feedback. How does this competition between replenishment by late-stage infall and depletion by external photoevaporation shape  disk populations?
\end{enumerate}

\noindent To understand these complicated environmental interactions, we require surveys of disks in more distant, more massive clusters, which are currently unfeasible with existing facilities.  ALMA2040, in connection with other upcomign facilities such as the ELT, will enable this. 


 \vspace{-0.4cm}
\section{Science Case for the 2040s}
 \vspace{-0.2cm}

ALMA2040 has the potential to impact many facets of the study of disks in clusters and their connection to environment, which is discussed more broadly in Section \ref{sec:widerProspects}. In this white paper we focus on a core science case,  the spirit of which is encapsulated in Figure \ref{fig:mainScience}. We base our understanding of the mass reservoir for planet formation and how it varies over time on samples of disks from low-mass stellar clusters within 0.4\,kpc. We also see some evidence for the ONC disks having lower masses, perhaps due to the harsher external radiation environment there. However, within 3\,kpc lies a much denser concentration of star-forming clouds, including much more massive clusters than previously explored, of the kind where most stars (and hence planets) are expected to form. Hence, our primary science goals are to:
 \begin{enumerate}
     \item Survey disk dust masses in all star-forming regions out to at least 3\,kpc down to a limiting mass of at least $0.5$\,M$_\oplus$, which would be sufficient to determine the slope and median of the disk mass distribution for Orion-like disk mass distributions (see Figure \ref{fig:SynthObs}). This will provide a complete sample of disk evolutionary phases across the stellar cluster mass function. 
    \item Determine how disk mass distributions depend upon age, stellar density, the radiation environment and surface density of the local interstellar medium.
 \end{enumerate}
Such capabilities will provide disk mass distributions similar to those in the left hand panel of Figure \ref{fig:mainScience} for a population of star-forming regions more than an order of magnitude more massive. For example, \cite{2025MNRAS.542...52Z}  find over 100 stellar cluster/molecular cloud associations within 3.5\,kpc. There are also around 10 well known clusters more massive than the ONC within 3\,kpc, including Trumpler 14, Trumpler 16, Cygnus OB2, RCW 38 and NGC 6611 with stellar cluster masses $>10^4$\,M$_\odot$. Note that many of these massive clusters are significantly closer than our 3\,kpc goal, such as those in Carina at $\sim2.3$\,kpc which would meet our requirements in 60\,per cent of the nominal observing time set by our requirements. The acquisition of such data would enable statistics on the planet forming reservoir across the cluster mass function.

\subsection{Wider science prospects for disks in clusters}
\label{sec:widerProspects}
An increase in sensitivity capable of  revolutionising our understanding of representative disks in distant massive star-forming clusters would also instigate a revolution in critical but hard-to-access tracers in disks in local lower mass star forming clusters. ALMA's capability to spatially and spectrally resolve gaseous disks, revealing chemistry and dynamics, has been showcased by several large programmes such as DSHARP \citep{2018ApJ...869L..41A}, AGE-PRO \citep{2025ApJ...989....1Z}, MAPS \citep{2021ApJS..257....1O}, DECO (Cleeves et al. submitted), exoALMA \citep{2025ApJ...984L...6T} and CHEER (Pegues et al. in prep). While such efforts are invaluable for tracing the thermodynamics and structure of the bulk component of protoplanetary disks, they are both time intensive and are incapable of high resolution detection of several molecular species that are critical to interpreting disk physics (particularly CO and its isotopologues), for example to measure disk gas masses. For example, MAPS studied the chemistry of 5 protoplanetary disks at 10\,au scales, whereas DECO studied 80 disks at scales a factor of 2-7 coarser. MAPS demonstrated that the chemical structure is complicated and coupled to the underlying gas/dust density distribution. DECO demonstrates the statistical variation as a function of star/disk/region parameters. However, with existing facilities it is prohibitively expensive to connect these contrasting approaches into a unified picture. 

Furthermore, even in nearby regions many disks remain too compact to be resolved by ALMA \citep[e.g.][]{2026A&A...710A.112R}.  Increasing ALMA sensitivity will push our understanding of fainter disks in nearby regions, particularly those around low mass stars that are the most common planet hosts. These advances will therefore revolutionise our understanding of the slope of the disk mass distribution as a function of time, stellar mass and  environment, a goal that is integral to understanding the origins of planetary systems.

There is recent renewed interest in the idea of infall onto disks, particularly beyond the Class 0/I phase. This is usually evidenced by elongated structures connecting to the disk referred to as ``streamers'' \citep[e.g.][]{2020NatAs...4.1158P, 2023EPJP..138..272K,  2024A&A...683A.133G}. However, the maximum recoverable scale of high-resolution ALMA observations means that these are usually filtered out and, when large scale structures are detected, gas kinematics is required to unambiguously conclude that they are associated with infall onto the disk. ALMA2040 would mostly provide impact here through its dynamic range that would currently require 2-3 ALMA configurations to replicate. It will therefore enable more efficient discovery and characterization and the ability to study the flows across scales that currently have to be probed with disparate observations.

\section{ALMA2040: Technical Requirements}

\subsection{Expectation based on scalings and prior observations }

\cite{2018ApJ...860...77E} surveyed the ONC ($d\sim0.39$\,kpc) in Band 7  with a 136-pointing mosaic, achieving a total on-source integration time of $\sim12$ hours. They reached a $4\sigma$ detection threshold of $\geq0.4$\,mJy ($\sim0.7$\,M$_\oplus$), and a median detected disk flux of $\sim0.5$\,mJy ($\sim0.9$\,M$_\oplus$). Notably, even at Orion's distance, this limiting mass already exceeds the 0.5\,M$_\oplus$ target of this white paper by a factor of $\sim2$, so reaching our science goal requires not just replicating that survey's depth at a much greater distance, but substantially improving on it. At the distance of Carina (2.3\,kpc), an equivalent ONC-like disk population is a factor ($2.3$\,kpc$ / 0.4$\,kpc)$^2$$\sim 33$ fainter, so reproducing the ONC survey with the same 12-hour integration would only reach a limiting mass of $\sim23$\,M$_\oplus$ -- a factor 46 above the 0.5\,M$_\oplus$ target. The proposed ALMA2040 factor $\sim$10 increase in sensitivity (stemming from increased collecting area and improved detectors), as well as the increase in bandwidth from 7.5 to $>$128\,GHz, will close this gap. Furthermore, the \cite{2018ApJ...860...77E} mosaic efficiency ($\sim1.65$ sources per pointing, spread over a comparatively low-density $1.5\arcmin\times1.5\arcmin$ field) is not representative of the much higher stellar densities in a cluster like Tr 14 in Carina. In such a cluster with typical source separations of $1–2\arcsec$, a single ALMA2040 pointing at $0.1\arcsec$ resolution will capture many more sources simultaneously. This additional efficiency gain is difficult to estimate reliably by analytic scaling alone, and so we proceed to simulated observations.


\subsection{Simulations with GHOST}
At a distance of around 2.3\,kpc, the Carina nebula hosts prototypical clusters that we would wish to target with future sub-mm observatories. For our exploration of synthetic observations we opt for an artificial cluster with a stellar density representative of that in the Carina cluster Trumpler 14 \citep[see e.g.][]{2024A&A...685A.100I}. Generating a cluster of disk models is computationally challenging and we do not expect to spatially resolve any disks. As a sensitivity exercise we therefore simulate a cluster of compact Gaussian sources with fluxes drawn from the distribution seen in the high UV environment of the Orion Nebula \citep{2018ApJ...860...77E}. We focus on the inner arcminute of our artificial cluster.

\begin{figure}
    \centering
    \includegraphics[width=\linewidth]{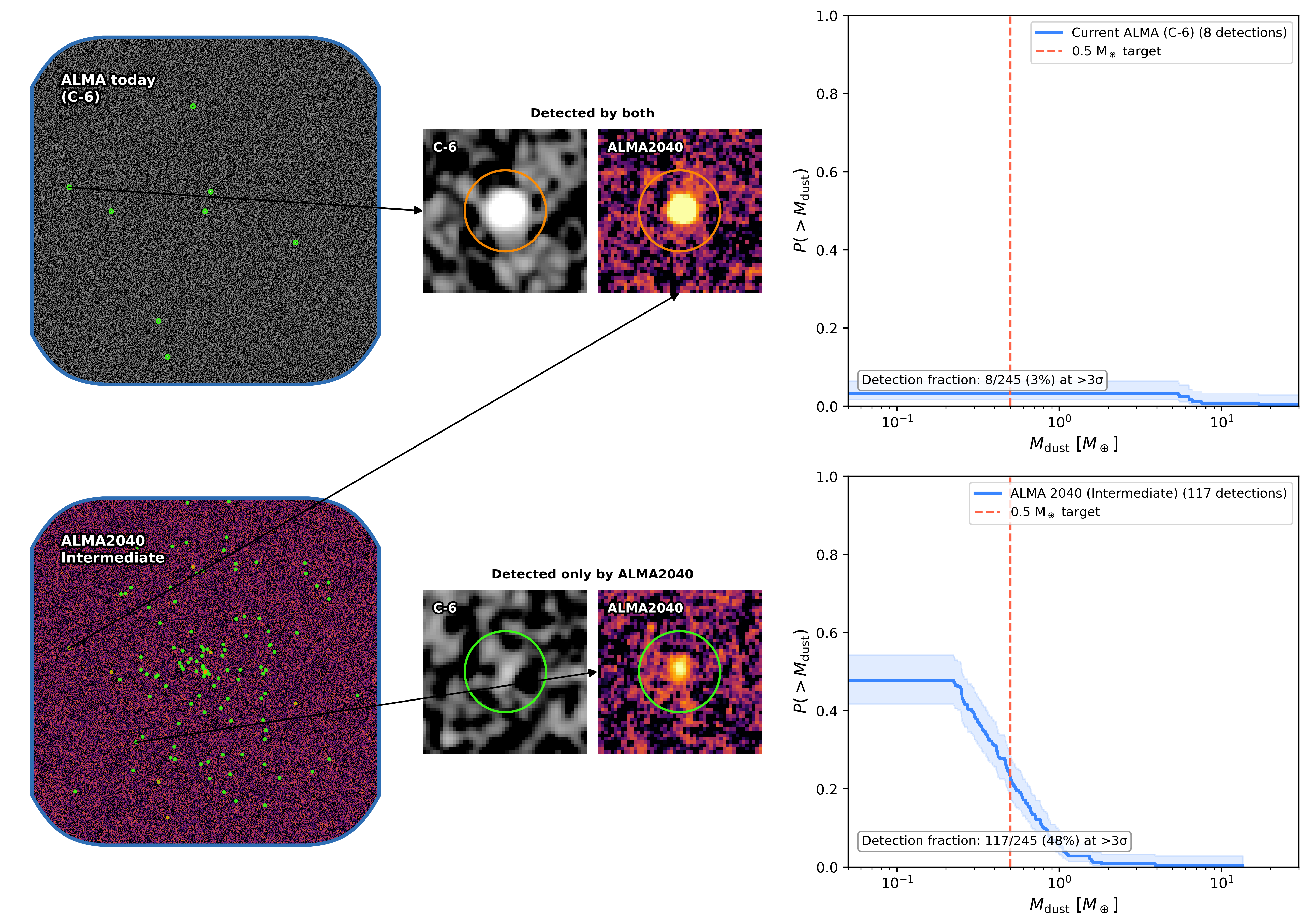}
    \caption{Synthetic observations of a cluster with ONC-like mass/luminosity distribution at a distance and stellar density representative of the Trumpler 14 in Carina. The left hand panels show the cleaned continuum images from ALMA C-6 and an example ALMA2040 configuration in the upper and lower panels respectively. To the right of those are zoom-ins on objects detected in both images (upper panels) and only by ALMA2040 (lower panels).  The right hand panels show the inferred disk mass distributions.  }
    \label{fig:SynthObs}
\end{figure}

We create synthetic continuum observations of this artificial cluster with the GHOST tools to compare ALMA C-6 and an example ALMA2040 configuration in current ALMA Band 7 (340\,GHz) mosaic observations with 12h on source in both cases. GHOST simulates a 7.5\,GHz bandwidth whereas ALMA2040 is expected to have a 128\,GHz bandwidth and so we account for this factor $\sim4$ increase in sensitivity relative to the raw GHOST output. The results of this exercise are shown in Figure \ref{fig:SynthObs}. The left hand panels show the continuum imaging over the same field, with the ALMA C-6 observations in the upper panel and the ALMA 2040 intermediate configuration in the lower panel, with detections highlighted by the green markers. To the right of these is a zoomed-in instance of a detection in both cases (upper set) and a detection only in the ALMA2040 case (lower set). The third set of panels show the cumulative disk mass distribution, similar to the upper left panel of Figure \ref{fig:mainScience}. This is computed using the standard optically thin emission approach 
	\begin{equation}
	    M_{\textrm{dust}} = \frac{D^2F_\nu}{\kappa_\nu B_\nu(T)}
	\end{equation}
where $D$ and $F_\nu$ are the target distance and flux, respectively, and at this point we assume a representative temperature $T=20\,K$ in the Planck function $B_
\nu(T)$ and an opacity $\kappa_\nu = 3.4$\,cm$^2$\,g$^{-1}$ at 890\,$\mu$m \citep[though note that the temperature may well be higher in strong radiation environments, e.g.][]{2021MNRAS.503.4172H}. 

The cleaned ALMA2040 data recovers 48\,\% of the injected objects in the image at $>3\sigma$ and is capable of reaching the 0.5\,M$_\oplus$ target mass within the assumed 12\,hours. Conversely, current ALMA C-6 detects only $<3$\,\% of targets in the same time.

This efficient observing is further motivated by the need to correct for free-free contamination from ionisation fronts in externally irradiated disks \citep{2023ApJ...954..127B}. The proposed multi-band nature of ALMA2040 would enable efficient and robust implementation of this correction.

\section{Synergies with current and future facilities}
\begin{figure}
    \centering
    \includegraphics[width=0.9\linewidth]{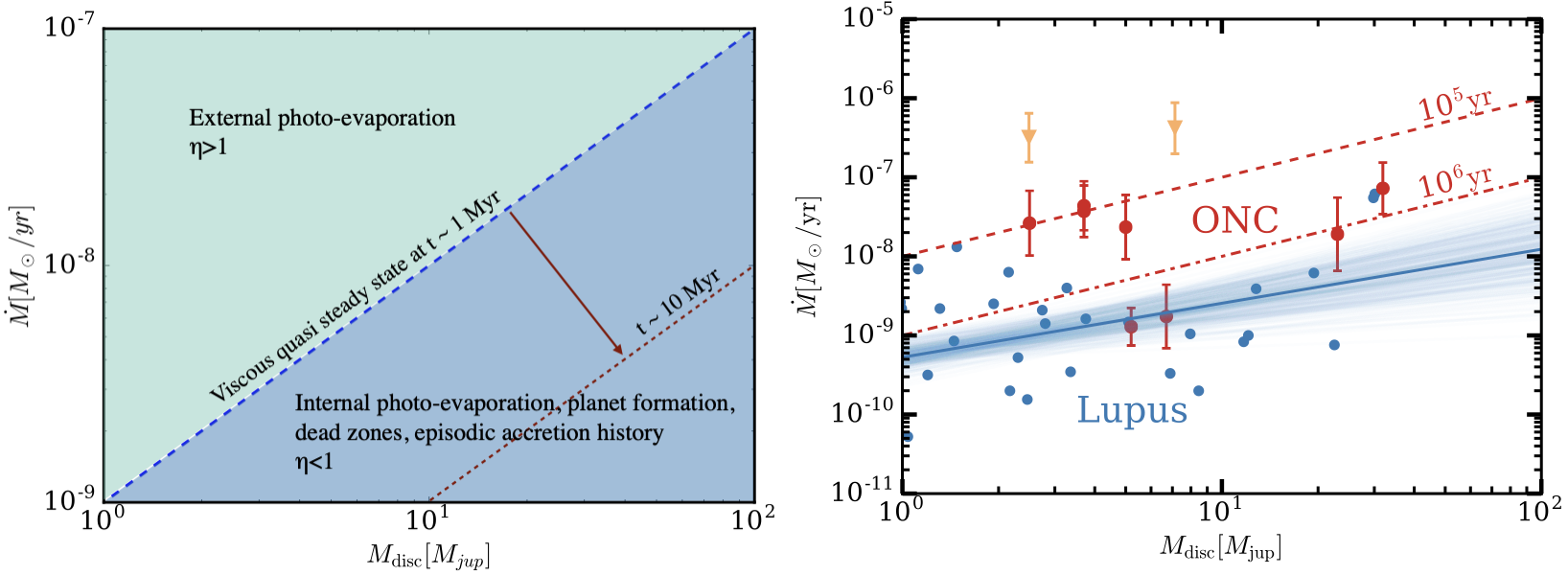}
    \caption{Existing and future surveys are estimating accretion rates in clusters out to 3\,kpc. ALMA2040 will provide the disk mass estimates that synergise with those accretion rates to distinguish disk evolutionary paradigms as a function of cluster mass and age \protect\citep{2017MNRAS.468.1631R}. Note that $M_\textrm{disc}$ in these plots is approximated as $100\times M_{\textrm{dust}}$. In the right hand panel orange points have suspected nebular contamination, making the accretion rate more uncertain \protect\citep[see][]{2017MNRAS.468.1631R}.}
    \label{fig:accretiontimescale}
\end{figure}

ALMA2040 observations of disks out to 3\,kpc would synergise strongly with existing and future observations that probe young stars and their disks in these clusters. For example, there are ongoing efforts to measure accretion rates in clusters within 3\,kpc including with VLT/MUSE \citep{2024A&A...685A.100I, 2025A&A...697A..39K} and the JWST NIRSpec MSA  \citep{2024A&A...684L...8R}.  These accretion rates, combined with the ALMA2040 disk mass constraints, provide an accretion timescale estimate. As illustrated in the left hand panel of Figure \ref{fig:accretiontimescale}, the accretion timescale is distinct depending on whether the disk is influenced by the environmental radiation field, evolving viscously in quasi-steady state, or evolving independently of the external radiation field in a non-viscous/non-steady fashion \citep{2017MNRAS.468.1631R}. The synergy of ALMA2040 with these facilities would hence enable a statistical diagnosis of how disks are evolving as a function of cluster mass and age. Hints of a sensitivity to environment are suggested based on the nearby clusters of Lupus and the ONC, as shown in the right hand panel of Figure \ref{fig:accretiontimescale}. The expectation is that the very massive clusters that could be probed by ALMA2040 would populate the upper-right region of this parameter space if external photoevaporation is a dominant driver of disk evolution in these regions.

An ALMA2040 campaign targeting distant stellar clusters would also synergise with facilities such as ngVLA, SKA and ELT to detect ionised gas associated with externally irradiated disks. For example, instruments like MICADO \citep{2024SPIE13096E..11S} and METIS \citep{2021Msngr.182...22B} will provide adaptive optics supported imaging and spectroscopy from 1.0 $\mu$m to 13 $\mu$m, achieving a similar spatial resolution ($\sim 20$ au) at 2-3 kpc as the current VLT in the ONC. Such observations will enable us to spatially resolve disks in distant clusters and provide multiple atomic lines (spectral resolving power from a few $\sim$ 1000 to 20,000) to characterize external photo-evaporation winds. In addition, the CO-rovibrational emission to which METIS is sensitive will be excited in the heated region at the base of the photoevaporative wind. These tracers are the hallmarks of strong UV fields, and can be correlated with sub-mm continuum-inferred disk masses.

\begin{tcolorbox}[boxsep=3pt,left=0pt,right=0pt,top=1pt,bottom=1pt,colframe=boxcolor,colback=boxcolor]
\justify
\textbf{Summary:} A future sub-mm multi-band observatory with an order of magnitude increase in sensitivity will enable the quantification of the distribution of disk continuum properties out to 3\,kpc from the Earth. This encapsulates very massive star forming regions unlike anything in the $\sim400$\,pc out to which disk properties are quantified at a population level to date, and is critical for determining the extent to which star-forming environments impact protoplanetary disk evolution. 
\end{tcolorbox}

\section*{Acknowledgements}
We thank the anonymous reviewer for their comments on the manuscript. 

This work was developed as part of the broader ALMA2040 community effort. We acknowledge the Lorentz Center and NOVA for supporting the community workshops and coordination activities that helped initiate and shape the ALMA2040 White Paper process.

T. J. Haworth  acknowledges a Royal Society Dorothy Hodgkin Fellowship and UKRI guaranteed funding for a Horizon Europe ERC consolidator grant (EP/Y024710/1). 
R. Boyden acknowledges support from the National Radio Astronomy Observatory Jansky Fellowship Program and NASA/Space Telescope Science Institute grants JWST-GO-06130 and JWST-GO-07534. 
AG acknowledges support from NSF AST-2407547 and the David and Lucile Packard Foundation and the Virginia Institute of Theoretical Astronomy (VITA). Á. Ribas has received funding from the Royal Society through a University Research Fellowship grant number URF\textbackslash R1\textbackslash 241791
A. J. Winter has been supported by the Royal Society through a University Research Fellowship grant number URF\textbackslash R1\textbackslash 241791.
T.~Birnstiel acknowledges funding from the European Union under the European Union's Horizon Europe Research and Innovation Programme 101124282 (EARLYBIRD) and funding by the Deutsche Forschungsgemeinschaft (DFG, German Research Foundation) under grant 325594231, and Germany's Excellence Strategy - EXC-2094 - 390783311. Views and opinions expressed are, however, those of the authors only and do not necessarily reflect those of the European Union or the European Research Council. Neither the European Union nor the granting authority can be held responsible for them.
D.Polychroni and D.Turrini acknowledge the support of the Italian Space Agency (ASI) through the ASI-INAF agreement no. 2021-5-HH.0 and addenda 2021-5-HH.1-2022 and 2021-5-HH.2- 2024. L.M. acknowledges funding support from the DAE through the NISER project RNI 4011. M.T. acknowledges funding from the European Research Council (ERC) via the ERC Advanced Grant PEBBLES (grant ID 101098309). 

\medskip

\section*{Affiliations}

\vspace{0.2em}
{
\footnotesize{
\noindent $^1$ School of Physical and Chemical Sciences, Queen Mary University of London, Mile End, London E1 4NS, UK, \\
$^2$ Dipartimento di Fisica “Aldo Pontremoli”, Universita degli
Studi di Milano, via Celoria 16, Milano, 20133, Italy, \\
$^3$ European Southern Observatory, Karl-Schwarzschild-Strasse 2, 85748 Garching bei München, Germany, \\
$^4$ School of Physics and Astronomy, University of Leicester, Leicester, LE1 7RH, United Kingdom, \\
$^5$ School of Physics, Trinity College Dublin, the University of Dublin, College Green, Dublin 2, Ireland, \\
$^6$ Ludwig-Maximilians-Universität München, Universitäts-Sternwarte, Scheinerstr. 1, 81679 München, Germany \& Exzellenzcluster ORIGINS, Boltzmannstr. 2, D-85748 Garching, Germany \\
$^{7}$ National Radio Astronomy Observatory, Charlottesville, VA 22903, USA,\\
$^8$ Department of Astronomy, University of Virginia, Charlottesville, VA 22904, USA \\
$^9$ Univ. Grenoble Alpes, CNRS, IPAG, 38000, Grenoble, France, \\
$^{10}$ Fakultät für Physik, Universität Duisburg-Essen, Lotharstraße 1, 47057 Duisburg, Germany, \\
$^{11}$  Mullard Space Science Laboratory, University College London, Holmbury St. Mary, Dorking, UK, \\
$^{12}$ Steward Observatory, University of Arizona, 933 N. Cherry
Ave, Tucson, AZ 85721-0065, USA, \\
$^{13}$ School of Physics and Astronomy, Monash University, Clayton VIC 3800, Australia, \\
$^{14}$ INAF - Osservatorio Astrofisico di Torino, Via Osservatorio 20, 10025, Pino Torinese (TO), Italy, \\
$^{15}$ Space Research Institute, Austrian Academy of Sciences,
Schmiedlstr. 6, A-8042, Graz, Austria, \\
$^{16}$ Departamento de Fisica aplicada III, ETSI Universidad de Sevilla, Camino de los Descubrimientos,
41092 Sevilla, \\
$^{17}$ Institut f{\"u}r Astro-und Teilchenphysik, Universit{\"a}t Innsbruck, Technikerstraße 25, 6020 Innsbruck, Austria, \\
$^{18}$ School of Physics and Astronomy, University of Leeds,
Leeds, United Kingdom, LS2 9JT \\
$^{19}$ Australia Telescope National Facility, CSIRO Space and Astronomy, PO Box 76, Epping, NSW 1710, Australia \\
$^{20}$ Konkoly Observatory, HUN-REN Research Centre for Astronomy and Earth Sciences, MTA Centre of Excellence, Konkoly-Thege Mikl\'os \'ut 15-17, 1121 Budapest, Hungary \\
$^{21}$ Department of Physics, Engineering Physics, and Astronomy
Queen's University, 64 Bader Lane, Kingston, Ontario, Canada, K7L 3N6 \\
$^{22}$ Institut de Ciències de l’Espai (ICE-CSIC), Campus UAB, Can Magrans S/N, 08193 Cerdanyola del Vallès, Catalonia, Spain \\
$^{23}$ Center for Astrophysics |Harvard \& Smithsonian, 60 Garden Street, Cambridge, MA 02138, USA\\
$^{24}$National Institute of Science Education and Research, Odisha, India\\
$^{25}$Homi Bhabha National Institute, Mumbai, India\\
$^{26}$Department of Astronomy, Stockholm University, Sweden\\
}
}

\bibliographystyle{apj_mod}
{\bibliography{bibliography_clusters}}

@ARTICLE{2016ApJ...831..125P,
       author = {{Pascucci}, I. and {Testi}, L. and {Herczeg}, G.~J. and {Long}, F. and {Manara}, C.~F. and {Hendler}, N. and {Mulders}, G.~D. and {Krijt}, S. and {Ciesla}, F. and {Henning}, Th. and {Mohanty}, S. and {Drabek-Maunder}, E. and {Apai}, D. and {Sz{\H{u}}cs}, L. and {Sacco}, G. and {Olofsson}, J.},
        title = "{A Steeper than Linear Disk Mass-Stellar Mass Scaling Relation}",
      journal = {\apj},
         year = 2016,
        month = nov,
       volume = {831},
       number = {2},
          eid = {125},
        pages = {125},
          doi = {10.3847/0004-637X/831/2/125},
archivePrefix = {arXiv},
       eprint = {1608.03621},
 primaryClass = {astro-ph.EP},
       adsurl = {https://ui.adsabs.harvard.edu/abs/2016ApJ...831..125P}
}

@ARTICLE{2025ApJ...984L...6T,
       author = {{Teague}, Richard and {Benisty}, Myriam and {Facchini}, Stefano and {Fukagawa}, Misato and {Pinte}, Christophe and {Andrews}, Sean M. and {Bae}, Jaehan and {Barraza-Alfaro}, Marcelo and {Cataldi}, Gianni and {Cuello}, Nicol{\'a}s and {Curone}, Pietro and {Czekala}, Ian and {Fasano}, Daniele and {Flock}, Mario and {Galloway-Sprietsma}, Maria and {Garg}, Himanshi and {Hall}, Cassandra and {Hammond}, Iain and {Hilder}, Thomas and {Huang}, Jane and {Ilee}, John D. and {Izquierdo}, Andr{\'e}s F. and {Kanagawa}, Kazuhiro and {Lesur}, Geoffroy and {Lodato}, Giuseppe and {Longarini}, Cristiano and {Loomis}, Ryan A. and {Masset}, Fr{\'e}d{\'e}ric and {Menard}, Francois and {Orihara}, Ryuta and {Price}, Daniel J. and {Rosotti}, Giovanni and {Stadler}, Jochen and {Testi}, Leonardo and {Yen}, Hsi-Wei and {Wafflard-Fernandez}, Gaylor and {Wilner}, David J. and {Winter}, Andrew J. and {W{\"o}lfer}, Lisa and {Yoshida}, Tomohiro C. and {Zawadzki}, Brianna},
        title = "{exoALMA. I. Science Goals, Project Design, and Data Products}",
      journal = {\apjl},
         year = 2025,
        month = may,
       volume = {984},
       number = {1},
          eid = {L6},
        pages = {L6},
          doi = {10.3847/2041-8213/adc43b},
archivePrefix = {arXiv},
       eprint = {2504.18688},
 primaryClass = {astro-ph.EP},
       adsurl = {https://ui.adsabs.harvard.edu/abs/2025ApJ...984L...6T}
}

@ARTICLE{2026A&A...710A.112R,
       author = {{Ricciardi}, G. and {Zagaria}, F. and {Miotello}, A. and {Manara}, C.~F. and {Rosotti}, G. and {Zallio}, L. and {Andrews}, S. and {Booth}, R. and {Carpenter}, J. and {Cleeves}, I. and et al.},
        title = "{Compact CO emission and no evidence of radial drift: ALMA observations of the faintest planet-forming disks in Lupus}",
      journal = {\aap},
         year = 2026,
        month = jun,
       volume = {710},
          eid = {A112},
        pages = {A112},
          doi = {10.1051/0004-6361/202558598},
archivePrefix = {arXiv},
       eprint = {2604.11292},
 primaryClass = {astro-ph.EP},
       adsurl = {https://ui.adsabs.harvard.edu/abs/2026A&A...710A.112R}
}

@ARTICLE{2021ApJS..257....1O,
       author = {{{\"O}berg}, Karin I. and {Guzm{\'a}n}, Viviana V. and {Walsh}, Catherine and {Aikawa}, Yuri and {Bergin}, Edwin A. and {Law}, Charles J. and {Loomis}, Ryan A. and {Alarc{\'o}n}, Felipe and {Andrews}, Sean M. and {Bae}, Jaehan and {Bergner}, Jennifer B. and {Boehler}, Yann and {Booth}, Alice S. and {Bosman}, Arthur D. and {Calahan}, Jenny K. and {Cataldi}, Gianni and {Cleeves}, L. Ilsedore and {Czekala}, Ian and {Furuya}, Kenji and {Huang}, Jane and {Ilee}, John D. and {Kurtovic}, Nicolas T. and {Le Gal}, Romane and {Liu}, Yao and {Long}, Feng and {M{\'e}nard}, Fran{\c{c}}ois and {Nomura}, Hideko and {P{\'e}rez}, Laura M. and {Qi}, Chunhua and {Schwarz}, Kamber R. and {Sierra}, Anibal and {Teague}, Richard and {Tsukagoshi}, Takashi and {Yamato}, Yoshihide and {van't Hoff}, Merel L.~R. and {Waggoner}, Abygail R. and {Wilner}, David J. and {Zhang}, Ke},
        title = "{Molecules with ALMA at Planet-forming Scales (MAPS). I. Program Overview and Highlights}",
      journal = {\apjs},
         year = 2021,
        month = nov,
       volume = {257},
       number = {1},
          eid = {1},
        pages = {1},
          doi = {10.3847/1538-4365/ac1432},
archivePrefix = {arXiv},
       eprint = {2109.06268},
 primaryClass = {astro-ph.EP},
       adsurl = {https://ui.adsabs.harvard.edu/abs/2021ApJS..257....1O}
}

@ARTICLE{2021MNRAS.503.4172H,
       author = {{Haworth}, Thomas J.},
        title = "{Warm millimetre dust in protoplanetary discs near massive stars}",
      journal = {\mnras},
         year = 2021,
        month = may,
       volume = {503},
       number = {3},
        pages = {4172-4182},
          doi = {10.1093/mnras/stab728},
archivePrefix = {arXiv},
       eprint = {2103.03950},
 primaryClass = {astro-ph.SR},
       adsurl = {https://ui.adsabs.harvard.edu/abs/2021MNRAS.503.4172H}
}

@ARTICLE{2026NatAs..10.1179P,
       author = {{Pedrini}, Alex and {Adamo}, Angela and {Calzetti}, Daniela and {Bik}, Arjan and {Haworth}, Thomas J. and {Elmegreen}, Bruce G. and {Krumholz}, Mark R. and {Linden}, Sean T. and {Gregg}, Benjamin and {Faustino Vieira}, Helena and {Bajaj}, Varun and {Ryon}, Jenna E. and {Ali}, Ahmad A. and {Andersson}, Eric P. and {Bortolini}, Giacomo and {Cignoni}, Michele and {Duarte-Cabral}, Ana and {Grasha}, Kathryn and {Lah{\'e}n}, Natalia and {Lai}, Thomas S.-Y. and {Lapeer}, Drew and {Messa}, Matteo and {{\"O}stlin}, G{\"o}ran and {Sabbi}, Elena and {Smith}, Linda J. and {Tosi}, Monica},
        title = "{The emerging timescale of young star clusters regulated by cluster stellar mass}",
      journal = {Nature Astronomy},
         year = 2026,
        month = aug,
       volume = {10},
        pages = {1179-1188},
          doi = {10.1038/s41550-026-02857-y},
archivePrefix = {arXiv},
       eprint = {2603.09867},
 primaryClass = {astro-ph.GA},
       adsurl = {https://ui.adsabs.harvard.edu/abs/2026NatAs..10.1179P}
}

@ARTICLE{2023EPJP..138..272K,
       author = {{Kuffmeier}, Michael and {Jensen}, Sigurd S. and {Haugb{\o}lle}, Troels},
        title = "{Rejuvenating infall: a crucial yet overlooked source of mass and angular momentum}",
      journal = {European Physical Journal Plus},
         year = 2023,
        month = mar,
       volume = {138},
       number = {3},
          eid = {272},
        pages = {272},
          doi = {10.1140/epjp/s13360-023-03880-y},
archivePrefix = {arXiv},
       eprint = {2303.05261},
 primaryClass = {astro-ph.SR},
       adsurl = {https://ui.adsabs.harvard.edu/abs/2023EPJP..138..272K}
}

@ARTICLE{2023ApJ...954..127B,
       author = {{Ballering}, Nicholas P. and {Cleeves}, L. Ilsedore and {Haworth}, Thomas J. and {Bally}, John and {Eisner}, Josh A. and {Ginsburg}, Adam and {Boyden}, Ryan D. and {Fang}, Min and {Kim}, Jinyoung Serena},
        title = "{Isolating Dust and Free-Free Emission in ONC Proplyds with ALMA Band 3 Observations}",
      journal = {\apj},
         year = 2023,
        month = sep,
       volume = {954},
       number = {2},
          eid = {127},
        pages = {127},
          doi = {10.3847/1538-4357/ace901},
archivePrefix = {arXiv},
       eprint = {2308.07369},
 primaryClass = {astro-ph.EP},
       adsurl = {https://ui.adsabs.harvard.edu/abs/2023ApJ...954..127B}
}

@ARTICLE{2025ApJ...989....1Z,
       author = {{Zhang}, Ke and {P{\'e}rez}, Laura M. and {Pascucci}, Ilaria and {Pinilla}, Paola and {Cieza}, Lucas A. and {Carpenter}, John and {Trapman}, Leon and {Deng}, Dingshan and {Agurto-Gangas}, Carolina and {Sierra}, Anibal and {Kurtovic}, Nicol{\'a}s T. and {Ruiz-Rodriguez}, Dary A. and {Vioque}, Miguel and {Miley}, James and {Tabone}, Beno{\^\i}t and {Gonz{\'a}lez-Ruilova}, Camilo and {Anania}, Rossella and {Rosotti}, Giovanni P. and {TorresVillanueva}, Estephani and {Hogerheijde}, Michiel R. and {Schwarz}, Kamber and {Kuznetsova}, Aleksandra},
        title = "{The ALMA Survey of Gas Evolution of PROtoplanetary Disks (AGE-PRO). I. Program Overview and Summary of First Results}",
      journal = {\apj},
         year = 2025,
        month = aug,
       volume = {989},
       number = {1},
          eid = {1},
        pages = {1},
          doi = {10.3847/1538-4357/addebe},
archivePrefix = {arXiv},
       eprint = {2506.10719},
 primaryClass = {astro-ph.EP},
       adsurl = {https://ui.adsabs.harvard.edu/abs/2025ApJ...989....1Z}
}

@ARTICLE{2018ApJ...869L..41A,
       author = {{Andrews}, Sean M. and {Huang}, Jane and {P{\'e}rez}, Laura M. and {Isella}, Andrea and {Dullemond}, Cornelis P. and {Kurtovic}, Nicol{\'a}s T. and {Guzm{\'a}n}, Viviana V. and {Carpenter}, John M. and {Wilner}, David J. and {Zhang}, Shangjia and {Zhu}, Zhaohuan and {Birnstiel}, Tilman and {Bai}, Xue-Ning and {Benisty}, Myriam and {Hughes}, A. Meredith and {{\"O}berg}, Karin I. and {Ricci}, Luca},
        title = "{The Disk Substructures at High Angular Resolution Project (DSHARP). I. Motivation, Sample, Calibration, and Overview}",
      journal = {\apjl},
         year = 2018,
        month = dec,
       volume = {869},
       number = {2},
          eid = {L41},
        pages = {L41},
          doi = {10.3847/2041-8213/aaf741},
archivePrefix = {arXiv},
       eprint = {1812.04040},
 primaryClass = {astro-ph.SR},
       adsurl = {https://ui.adsabs.harvard.edu/abs/2018ApJ...869L..41A}
}

@ARTICLE{2025MNRAS.542...52Z,
       author = {{Zhou}, Ji-Xuan and {Li}, Guang-Xing and {Chen}, Bing-Qiu},
        title = "{Kinematics of the interstellar medium using Gaia: A catalogue of 102 young stellar object─molecular cloud associations within 3.5 kpc of the Sun with 3D velocities}",
      journal = {\mnras},
         year = 2025,
        month = sep,
       volume = {542},
       number = {1},
        pages = {52-61},
          doi = {10.1093/mnras/staf1222},
       adsurl = {https://ui.adsabs.harvard.edu/abs/2025MNRAS.542...52Z}
}

@ARTICLE{2016A&A...595A...1G,
       author = {{Gaia Collaboration} and {Prusti}, T. and {de Bruijne}, J.~H.~J. and {Brown}, A.~G.~A. and {Vallenari}, A. and {Babusiaux}, C. and {Bailer-Jones}, C.~A.~L. and {Bastian}, U. and {Biermann}, M. and {Evans}, D.~W. and {Eyer}, L. and {Jansen}, F. and {Jordi}, C. and {Klioner}, S.~A. and {Lammers}, U. and {Lindegren}, L. and {Luri}, X. and {Mignard}, F. and {Milligan}, D.~J. and {Panem}, C. and {Poinsignon}, V. and {Pourbaix}, D. and {Randich}, S. and {Sarri}, G. and {Sartoretti}, P. and {Siddiqui}, H.~I. and {Soubiran}, C. and {Valette}, V. and {van Leeuwen}, F. and {Walton}, N.~A. and {Aerts}, C. and {Arenou}, F. and {Cropper}, M. and {Drimmel}, R. and {H{\o}g}, E. and {Katz}, D. and {Lattanzi}, M.~G. and {O'Mullane}, W. and {Grebel}, E.~K. and {Holland}, A.~D. and {Huc}, C. and {Passot}, X. and {Bramante}, L. and {Cacciari}, C. and {Casta{\~n}eda}, J. and {Chaoul}, L. and {Cheek}, N. and {De Angeli}, F. and {Fabricius}, C. and {Guerra}, R. and {Hern{\'a}ndez}, J. and {Jean-Antoine-Piccolo}, A. and {Masana}, E. and {Messineo}, R. and {Mowlavi}, N. and {Nienartowicz}, K. and {Ord{\'o}{\~n}ez-Blanco}, D. and {Panuzzo}, P. and {Portell}, J. and {Richards}, P.~J. and {Riello}, M. and {Seabroke}, G.~M. and {Tanga}, P. and {Th{\'e}venin}, F. and {Torra}, J. and {Els}, S.~G. and {Gracia-Abril}, G. and {Comoretto}, G. and {Garcia-Reinaldos}, M. and {Lock}, T. and {Mercier}, E. and {Altmann}, M. and {Andrae}, R. and {Astraatmadja}, T.~L. and {Bellas-Velidis}, I. and {Benson}, K. and {Berthier}, J. and {Blomme}, R. and {Busso}, G. and {Carry}, B. and {Cellino}, A. and {Clementini}, G. and {Cowell}, S. and {Creevey}, O. and {Cuypers}, J. and {Davidson}, M. and {De Ridder}, J. and {de Torres}, A. and {Delchambre}, L. and {Dell'Oro}, A. and {Ducourant}, C. and {Fr{\'e}mat}, Y. and {Garc{\'\i}a-Torres}, M. and {Gosset}, E. and {Halbwachs}, J.-L. and {Hambly}, N.~C. and {Harrison}, D.~L. and {Hauser}, M. and {Hestroffer}, D. and {Hodgkin}, S.~T. and {Huckle}, H.~E. and {Hutton}, A. and {Jasniewicz}, G. and {Jordan}, S. and {Kontizas}, M. and {Korn}, A.~J. and {Lanzafame}, A.~C. and {Manteiga}, M. and {Moitinho}, A. and {Muinonen}, K. and {Osinde}, J. and {Pancino}, E. and {Pauwels}, T. and {Petit}, J.-M. and {Recio-Blanco}, A. and {Robin}, A.~C. and {Sarro}, L.~M. and {Siopis}, C. and {Smith}, M. and {Smith}, K.~W. and {Sozzetti}, A. and {Thuillot}, W. and {van Reeven}, W. and {Viala}, Y. and {Abbas}, U. and {Abreu Aramburu}, A. and {Accart}, S. and {Aguado}, J.~J. and {Allan}, P.~M. and {Allasia}, W. and {Altavilla}, G. and {{\'A}lvarez}, M.~A. and {Alves}, J. and {Anderson}, R.~I. and {Andrei}, A.~H. and {Anglada Varela}, E. and {Antiche}, E. and {Antoja}, T. and {Ant{\'o}n}, S. and {Arcay}, B. and {Atzei}, A. and {Ayache}, L. and {Bach}, N. and {Baker}, S.~G. and {Balaguer-N{\'u}{\~n}ez}, L. and {Barache}, C. and {Barata}, C. and {Barbier}, A. and {Barblan}, F. and {Baroni}, M. and {Barrado y Navascu{\'e}s}, D. and {Barros}, M. and {Barstow}, M.~A. and {Becciani}, U. and {Bellazzini}, M. and {Bellei}, G. and {Bello Garc{\'\i}a}, A. and {Belokurov}, V. and {Bendjoya}, P. and {Berihuete}, A. and {Bianchi}, L. and {Bienaym{\'e}}, O. and {Billebaud}, F. and {Blagorodnova}, N. and {Blanco-Cuaresma}, S. and {Boch}, T. and {Bombrun}, A. and {Borrachero}, R. and {Bouquillon}, S. and {Bourda}, G. and {Bouy}, H. and {Bragaglia}, A. and {Breddels}, M.~A. and {Brouillet}, N. and {Br{\"u}semeister}, T. and {Bucciarelli}, B. and {Budnik}, F. and {Burgess}, P. and {Burgon}, R. and {Burlacu}, A. and {Busonero}, D. and {Buzzi}, R. and {Caffau}, E. and {Cambras}, J. and {Campbell}, H. and {Cancelliere}, R. and {Cantat-Gaudin}, T. and {Carlucci}, T. and {Carrasco}, J.~M. and {Castellani}, M. and {Charlot}, P. and {Charnas}, J. and {Charvet}, P. and {Chassat}, F. and {Chiavassa}, A. and {Clotet}, M. and {Cocozza}, G. and {Collins}, R.~S. and {Collins}, P. and {Costigan}, G.},
        title = "{The Gaia mission}",
      journal = {\aap},
         year = 2016,
        month = nov,
       volume = {595},
          eid = {A1},
        pages = {A1},
          doi = {10.1051/0004-6361/201629272},
archivePrefix = {arXiv},
       eprint = {1609.04153},
 primaryClass = {astro-ph.IM},
       adsurl = {https://ui.adsabs.harvard.edu/abs/2016A&A...595A...1G}
}

@ARTICLE{2018ApJ...860...77E,
       author = {{Eisner}, J.~A. and {Arce}, H.~G. and {Ballering}, N.~P. and {Bally}, J. and {Andrews}, S.~M. and {Boyden}, R.~D. and {Di Francesco}, J. and {Fang}, M. and {Johnstone}, D. and {Kim}, J.~S. and {Mann}, R.~K. and {Matthews}, B. and {Pascucci}, I. and {Ricci}, L. and {Sheehan}, P.~D. and {Williams}, J.~P.},
        title = "{Protoplanetary Disk Properties in the Orion Nebula Cluster: Initial Results from Deep, High-resolution ALMA Observations}",
      journal = {\apj},
         year = 2018,
        month = jun,
       volume = {860},
       number = {1},
          eid = {77},
        pages = {77},
          doi = {10.3847/1538-4357/aac3e2},
archivePrefix = {arXiv},
       eprint = {1805.03669},
 primaryClass = {astro-ph.SR},
       adsurl = {https://ui.adsabs.harvard.edu/abs/2018ApJ...860...77E}
}

@INPROCEEDINGS{2023ASPC..534..539M,
       author = {{Manara}, C.~F. and {Ansdell}, M. and {Rosotti}, G.~P. and {Hughes}, A.~M. and {Armitage}, P.~J. and {Lodato}, G. and {Williams}, J.~P.},
        title = "{Demographics of Young Stars and their Protoplanetary Disks: Lessons Learned on Disk Evolution and its Connection to Planet Formation}",
    booktitle = {Protostars and Planets VII},
         year = 2023,
       editor = {{Inutsuka}, S. and {Aikawa}, Y. and {Muto}, T. and {Tomida}, K. and {Tamura}, M.},
       series = {Astronomical Society of the Pacific Conference Series},
       volume = {534},
        month = jul,
        pages = {539},
          doi = {10.48550/arXiv.2203.09930},
archivePrefix = {arXiv},
       eprint = {2203.09930},
 primaryClass = {astro-ph.SR},
       adsurl = {https://ui.adsabs.harvard.edu/abs/2023ASPC..534..539M}
}

@ARTICLE{2021Msngr.182...22B,
       author = {{Brandl}, B. and {Bettonvil}, F. and {van Boekel}, R. and {Glauser}, A. and {Quanz}, S. and {Absil}, O. and {Amorim}, A. and {Feldt}, M. and {Glasse}, A. and {G{\"u}del}, M. and {Ho}, P. and {Labadie}, L. and {Meyer}, M. and {Pantin}, E. and {van Winckel}, H. and {METIS Consortium}},
        title = "{METIS: The Mid-infrared ELT Imager and Spectrograph}",
      journal = {The Messenger},
         year = 2021,
        month = mar,
       volume = {182},
        pages = {22-26},
          doi = {10.18727/0722-6691/5218},
archivePrefix = {arXiv},
       eprint = {2103.11208},
 primaryClass = {astro-ph.IM},
       adsurl = {https://ui.adsabs.harvard.edu/abs/2021Msngr.182...22B}
}

@INPROCEEDINGS{2024SPIE13096E..11S,
       author = {{Sturm}, E. and {Davies}, R. and {Alves}, J. and {Cl{\'e}net}, Y. and {Kotilainen}, J. and {Monna}, A. and {Nicklas}, H. and {Pott}, J.-U. and {Tolstoy}, E. and {Vulcani}, B. and {Achren}, J. and {Annadevara}, S. and {Anwand-Heerwart}, H. and {Arcidiacono}, C. and {Barboza}, S. and {Barl}, L. and {Baudoz}, P. and {Bender}, R. and {Bezawada}, N. and {Biondi}, F. and {Bizenberger}, P. and {Blin}, A. and {Bon{\'e}}, A. and {Bonifacio}, P. and {Borgo}, B. and {Born}, J. van den and {Buey}, T. and {Cao}, Y. and {Chapron}, F. and {Chauvin}, G. and {Chemla}, F. and {Cloiseau}, K. and {Cohen}, M. and {Colin}, C. and {Czoske}, O. and {Dette}, J.-O. and {Deysenroth}, M. and {Dijkstra}, E. and {Dreizler}, S. and {Dupuis}, O. and {Egmond}, G. van and {Eisenhauer}, F. and {Elswijk}, E. and {Emslander}, A. and {Fabricius}, M. and {Fasola}, G. and {Ferreira}, F. and {F{\"o}rster Schreiber}, N. and {Fontana}, A. and {Gaudemard}, J. and {Gautherot}, N. and {Gendron}, E. and {Gennet}, C. and {Genzel}, R. and {Ghouchou}, L. and {Gillessen}, S. and {Gratadour}, D. and {Grazian}, A. and {Grupp}, F. and {Guieu}, S. and {Gullieuszik}, M. and {Haan}, M. de and {Hartke}, J. and {Hartl}, M. and {Haussmann}, F. and {Helin}, T. and {Hess}, H.-J. and {Hofferbert}, R. and {Huber}, H. and {Huby}, E. and {Huet}, J.-M. and {Ives}, D. and {Janssen}, A. and {Jaufmann}, P. and {Jilg}, T. and {Jodlbauer}, D. and {Jost}, J. and {Kausch}, W. and {Kellermann}, H. and {Kerber}, F. and {Kravcar}, H. and {Kravchenko}, K. and {Kulcs{\'a}r}, C. and {Kuncarayakti}, H. and {Kunst}, P. and {Kwast}, S. and {Lang}, F. and {Lange}, J. and {Lapeyrere}, V. and {Le Ruyet}, B. and {Leschinski}, K. and {Locatelli}, H. and {Massari}, D. and {Mattila}, S. and {Mei}, S. and {Merlin}, F. and {Meyer}, E. and {Michel}, C. and {Mohr}, L. and {Montarg{\`e}s}, M. and {M{\"u}ller}, F. and {M{\"u}nch}, N. and {Navarro}, R. and {Neumann}, U. and {Neumayer}, N. and {Neumeier}, L. and {Pedichini}, F. and {Pfl{\"u}ger}, A. and {Piazzesi}, R. and {Pinard}, L. and {Porras}, J. and {Portulari}, E. and {Przybilla}, N. and {Rabien}, S. and {Raffard}, J. and {Ragazzoni}, R. and {Ramlau}, R. and {Ramos}, J. and {Ramsay}, S. and {Raynaud}, H.-F. and {Rhode}, P. and {Richter}, A. and {Rix}, H.-W. and {Rodenhuis}, M. and {Rohloff}, R.-R. and {Romp}, R. and {Rousselot}, P. and {Sabha}, N. and {Sassolas}, B. and {Schlichter}, J. and {Schuil}, M. and {Schweitzer}, M. and {Seemann}, U. and {Sevin}, A. and {Simioni}, M. and {Spallek}, L. and {S{\"o}nmez}, A. and {Suuronen}, J. and {Taburet}, S. and {Thomas}, J. and {Tisserand}, E. and {Vaccari}, P. and {Valenti}, E. and {Verdoes Kleijn}, G. and {Verdugo}, M. and {Vidal}, F. and {Wagner}, R. and {Wegner}, M. and {Winden}, D. van and {Witschel}, J. and {Zanella}, A. and {Zeilinger}, W. and {Ziegleder}, J. and {Ziegler}, B.},
        title = "{The MICADO first light imager for the ELT: overview and current status}",
    booktitle = {Ground-based and Airborne Instrumentation for Astronomy X},
         year = 2024,
       editor = {{Bryant}, Julia J. and {Motohara}, Kentaro and {Vernet}, Jo{\"e}l. R.~D.},
       series = {Society of Photo-Optical Instrumentation Engineers (SPIE) Conference Series},
       volume = {13096},
        month = jul,
          eid = {1309611},
        pages = {1309611},
          doi = {10.1117/12.3017752},
archivePrefix = {arXiv},
       eprint = {2408.16396},
 primaryClass = {astro-ph.IM},
       adsurl = {https://ui.adsabs.harvard.edu/abs/2024SPIE13096E..11S}
}

@ARTICLE{2025A&A...697A..39K,
       author = {{Kang}, Da Eun and {Itrich}, Dominika and {Ksoll}, Victor F. and {Testi}, Leonardo and {Klessen}, Ralf S. and {Molinari}, Sergio},
        title = "{Spectral classification of young stars using conditional invertible neural networks: II. Application to Trumpler 14 in Carina}",
      journal = {\aap},
         year = 2025,
        month = may,
       volume = {697},
          eid = {A39},
        pages = {A39},
          doi = {10.1051/0004-6361/202450394},
archivePrefix = {arXiv},
       eprint = {2503.19697},
 primaryClass = {astro-ph.SR},
       adsurl = {https://ui.adsabs.harvard.edu/abs/2025A&A...697A..39K}
}

@ARTICLE{2024A&A...685A.100I,
       author = {{Itrich}, Dominika and {Testi}, Leonardo and {Beccari}, Giacomo and {Manara}, Carlo F. and {Reiter}, Megan and {Preibisch}, Thomas and {McLeod}, Anna F. and {Rosotti}, Giovanni and {Klessen}, Ralf and {Molinari}, Sergio and {Hennebelle}, Patrick},
        title = "{The population of young low-mass stars in Trumpler 14}",
      journal = {\aap},
         year = 2024,
        month = may,
       volume = {685},
          eid = {A100},
        pages = {A100},
          doi = {10.1051/0004-6361/202347380},
archivePrefix = {arXiv},
       eprint = {2309.14168},
 primaryClass = {astro-ph.SR},
       adsurl = {https://ui.adsabs.harvard.edu/abs/2024A&A...685A.100I}
}

@ARTICLE{2024A&A...684L...8R,
       author = {{Rogers}, Ciar{\'a}n and {de Marchi}, Guido and {Brandl}, Bernhard},
        title = "{Determining stellar accretion rates from Pa$_{{\ensuremath{\alpha}}}$ and Br$_{{\ensuremath{\beta}}}$ emission lines with JWST NIRSpec. Accretion of pre-main-sequence stars in NGC 3603}",
      journal = {\aap},
         year = 2024,
        month = apr,
       volume = {684},
          eid = {L8},
        pages = {L8},
          doi = {10.1051/0004-6361/202449282},
archivePrefix = {arXiv},
       eprint = {2403.09568},
 primaryClass = {astro-ph.SR},
       adsurl = {https://ui.adsabs.harvard.edu/abs/2024A&A...684L...8R}
}

@ARTICLE{2026MNRAS.546ag034Q,
       author = {{Qiao}, Lin and {Coleman}, Gavin A.~L. and {Haworth}, Thomas J.},
        title = "{Formation of multiplanetary systems via pebble accretion in externally photoevaporating discs in stellar clusters}",
      journal = {\mnras},
         year = 2026,
        month = mar,
       volume = {546},
       number = {3},
          eid = {stag034},
        pages = {stag034},
          doi = {10.1093/mnras/stag034},
archivePrefix = {arXiv},
       eprint = {2601.03963},
 primaryClass = {astro-ph.EP},
       adsurl = {https://ui.adsabs.harvard.edu/abs/2026MNRAS.546ag034Q}
}

@ARTICLE{2016ApJ...828...46A,
       author = {{Ansdell}, M. and {Williams}, J.~P. and {van der Marel}, N. and {Carpenter}, J.~M. and {Guidi}, G. and {Hogerheijde}, M. and {Mathews}, G.~S. and {Manara}, C.~F. and {Miotello}, A. and {Natta}, A. and {Oliveira}, I. and {Tazzari}, M. and {Testi}, L. and {van Dishoeck}, E.~F. and {van Terwisga}, S.~E.},
        title = "{ALMA Survey of Lupus Protoplanetary Disks. I. Dust and Gas Masses}",
      journal = {\apj},
         year = 2016,
        month = sep,
       volume = {828},
       number = {1},
          eid = {46},
        pages = {46},
          doi = {10.3847/0004-637X/828/1/46},
archivePrefix = {arXiv},
       eprint = {1604.05719},
 primaryClass = {astro-ph.EP},
       adsurl = {https://ui.adsabs.harvard.edu/abs/2016ApJ...828...46A}
}

@ARTICLE{2017AJ....153..240A,
       author = {{Ansdell}, M. and {Williams}, J.~P. and {Manara}, C.~F. and {Miotello}, A. and {Facchini}, S. and {van der Marel}, N. and {Testi}, L. and {van Dishoeck}, E.~F.},
        title = "{An ALMA Survey of Protoplanetary Disks in the {\ensuremath{\sigma}} Orionis Cluster}",
      journal = {\aj},
         year = 2017,
        month = may,
       volume = {153},
       number = {5},
          eid = {240},
        pages = {240},
          doi = {10.3847/1538-3881/aa69c0},
archivePrefix = {arXiv},
       eprint = {1703.08546},
 primaryClass = {astro-ph.EP},
       adsurl = {https://ui.adsabs.harvard.edu/abs/2017AJ....153..240A}
}

@ARTICLE{2024A&A...683A.133G,
       author = {{Gupta}, Aashish and {Miotello}, Anna and {Williams}, Jonathan P. and {Birnstiel}, Til and {Kuffmeier}, Michael and {Yen}, Hsi-Wei},
        title = "{TIPSY: Trajectory of Infalling Particles in Streamers around Young stars. Dynamical analysis of the streamers around S CrA and HL Tau}",
      journal = {\aap},
         year = 2024,
        month = mar,
       volume = {683},
          eid = {A133},
        pages = {A133},
          doi = {10.1051/0004-6361/202348007},
archivePrefix = {arXiv},
       eprint = {2401.10403},
 primaryClass = {astro-ph.SR},
       adsurl = {https://ui.adsabs.harvard.edu/abs/2024A&A...683A.133G}
}

@ARTICLE{2020NatAs...4.1158P,
       author = {{Pineda}, Jaime E. and {Segura-Cox}, Dominique and {Caselli}, Paola and {Cunningham}, Nichol and {Zhao}, Bo and {Schmiedeke}, Anika and {Maureira}, Mar{\'\i}a Jos{\'e} and {Neri}, Roberto},
        title = "{A protostellar system fed by a streamer of 10,500 au length}",
      journal = {Nature Astronomy},
         year = 2020,
        month = jan,
       volume = {4},
        pages = {1158-1163},
          doi = {10.1038/s41550-020-1150-z},
archivePrefix = {arXiv},
       eprint = {2007.13430},
 primaryClass = {astro-ph.GA},
       adsurl = {https://ui.adsabs.harvard.edu/abs/2020NatAs...4.1158P}
}

@ARTICLE{2025OJAp....8E..54A,
       author = {{Allen}, Megan and {Anania}, Rossella and {Andersen}, Morten and {Aru}, Mari-Liis and {Ballabio}, Giulia and {Ballering}, Nicholas P. and {Beccari}, Giacomo and {Bern{\'e}}, Olivier and {Bik}, Arjan and {Boyden}, Ryan and {Coleman}, Gavin and {D{\'\i}az-Berrios}, Javiera and {Eatson}, Joseph W. and {Frediani}, Jenny and {Forbrich}, Jan and {Gkimisi}, Katia and {Goicoechea}, Javier R. and {Gupta}, Saumya and {Guarcello}, Mario G. and {Haworth}, Thomas J. and {Henney}, William J. and {Isella}, Andrea and {Itrich}, Dominika and {Keyte}, Luke and {Kim}, Jinyoung Serena and {Kuhn}, Michael and {Le Petit}, Frank and {Luo}, Lilian and {Manara}, Carlo and {Mauco}, Karina and {Meshaka}, Rapha{\"e}l and {Millstone}, Samuel and {Owen}, James E. and {Paine}, S{\'e}bastien and {Parker}, Richard and {Peake}, Tyger and {Peatt}, Megan and {Pinilla}, Paola and {Qiao}, Lin and {Ram{\'\i}rez-Tannus}, Mar{\'\i}a Claudia and {Ramsay}, Suzanne and {Reiter}, Megan and {Rogers}, Ciar{\'a}n and {Rosotti}, Giovanni and {Schroetter}, Ilane and {Sellek}, Andrew and {Testi}, Leonardo and {van Terwisga}, Sierk and {Vicente}, Silvia and {Walsh}, Catherine and {Winter}, Andrew and {Wright}, Nicholas J. and {Zeidler}, Peter},
        title = "{The past, present and future of observations of externally irradiated disks}",
      journal = {The Open Journal of Astrophysics},
         year = 2025,
        month = may,
       volume = {8},
          eid = {54},
        pages = {54},
          doi = {10.33232/001c.137538},
archivePrefix = {arXiv},
       eprint = {2502.12255},
 primaryClass = {astro-ph.SR},
       adsurl = {https://ui.adsabs.harvard.edu/abs/2025OJAp....8E..54A}
}

@ARTICLE{2022EPJP..137.1132W,
       author = {{Winter}, Andrew J. and {Haworth}, Thomas J.},
        title = "{The external photoevaporation of planet-forming discs}",
      journal = {European Physical Journal Plus},
         year = 2022,
        month = oct,
       volume = {137},
       number = {10},
          eid = {1132},
        pages = {1132},
          doi = {10.1140/epjp/s13360-022-03314-1},
archivePrefix = {arXiv},
       eprint = {2206.11910},
 primaryClass = {astro-ph.EP},
       adsurl = {https://ui.adsabs.harvard.edu/abs/2022EPJP..137.1132W}
}

@ARTICLE{2023EPJP..138...11C,
       author = {{Cuello}, Nicol{\'a}s and {M{\'e}nard}, Fran{\c{c}}ois and {Price}, Daniel J.},
        title = "{Close encounters: How stellar flybys shape planet-forming discs}",
      journal = {European Physical Journal Plus},
         year = 2023,
        month = jan,
       volume = {138},
       number = {1},
          eid = {11},
        pages = {11},
          doi = {10.1140/epjp/s13360-022-03602-w},
archivePrefix = {arXiv},
       eprint = {2207.09752},
 primaryClass = {astro-ph.EP},
       adsurl = {https://ui.adsabs.harvard.edu/abs/2023EPJP..138...11C}
}

@ARTICLE{2025ApJ...979..120H,
       author = {{Hallatt}, Tim and {Lee}, Eve J.},
        title = "{On the Formation of Planets in the Milky Way's Thick Disk}",
      journal = {\apj},
         year = 2025,
        month = feb,
       volume = {979},
       number = {2},
          eid = {120},
        pages = {120},
          doi = {10.3847/1538-4357/ad9aa1},
archivePrefix = {arXiv},
       eprint = {2408.09319},
 primaryClass = {astro-ph.EP},
       adsurl = {https://ui.adsabs.harvard.edu/abs/2025ApJ...979..120H}
}

@ARTICLE{2023MNRAS.522.1939Q,
       author = {{Qiao}, Lin and {Coleman}, Gavin A.~L. and {Haworth}, Thomas J.},
        title = "{Planet formation via pebble accretion in externally photoevaporating discs}",
      journal = {\mnras},
         year = 2023,
        month = jun,
       volume = {522},
       number = {2},
        pages = {1939-1950},
          doi = {10.1093/mnras/stad944},
archivePrefix = {arXiv},
       eprint = {2303.15177},
 primaryClass = {astro-ph.EP},
       adsurl = {https://ui.adsabs.harvard.edu/abs/2023MNRAS.522.1939Q}
}

@ARTICLE{2017MNRAS.468.1631R,
       author = {{Rosotti}, Giovanni P. and {Clarke}, Cathie J. and {Manara}, Carlo F. and {Facchini}, Stefano},
        title = "{Constraining protoplanetary disc evolution using accretion rate and disc mass measurements: the usefulness of the dimensionless accretion parameter}",
      journal = {\mnras},
         year = 2017,
        month = jun,
       volume = {468},
       number = {2},
        pages = {1631-1638},
          doi = {10.1093/mnras/stx595},
archivePrefix = {arXiv},
       eprint = {1703.02974},
 primaryClass = {astro-ph.SR},
       adsurl = {https://ui.adsabs.harvard.edu/abs/2017MNRAS.468.1631R}
}

@ARTICLE{2025arXiv250509754V,
       author = {{Valencia}, Diana and {Moro-Martin}, Amaya and {Teske}, Johanna},
        title = "{Diversity of Exoplanets}",
      journal = {arXiv e-prints},
         year = 2025,
        month = may,
          eid = {arXiv:2505.09754},
        pages = {arXiv:2505.09754},
          doi = {10.48550/arXiv.2505.09754},
archivePrefix = {arXiv},
       eprint = {2505.09754},
 primaryClass = {astro-ph.EP},
       adsurl = {https://ui.adsabs.harvard.edu/abs/2025arXiv250509754V}
}
\end{document}